\documentclass[sigconf]{acmart}

\usepackage{balance}
\newcommand{\toolname}{GAPP}
\newcommand{\cm}{{\em CMetric}}
\newcommand{\subsub}[1]{\subsubsection*{\bf #1}}

\makeatletter
\newcommand\notsotiny{\@setfontsize\notsotiny\@vipt\@viipt}
\makeatother

\usepackage{tikz}
\usetikzlibrary{patterns}
\usepackage{pgfplots}
\usepackage{pgfplotstable}
\pgfplotsset{compat=1.14}
\usepackage{graphicx}
\usepackage{graphics}
\usepackage{wrapfig}

\def\BibTeX{{\rm B\kern-.05em{\sc i\kern-.025em b}\kern-.08emT\kern-.1667em\lower.7ex\hbox{E}\kern-.125emX}}
    
\copyrightyear{2020}
\acmYear{2020}
\setcopyright{acmcopyright}
\acmConference[ICPE '20]{Proceedings of the 2020 ACM/SPEC International Conference on Performance Engineering}{April 20--24, 2020}{Edmonton, AB, Canada} 
\acmBooktitle{Proceedings of the 2020 ACM/SPEC International Conference on Performance Engineering (ICPE '20), April 20--24, 2020, Edmonton, AB, Canada} 
\acmPrice{15.00}
\acmDOI{10.1145/XXXXXX.XXXXXX}
\acmISBN{978-1-4503-6286-3/19/04} 

\copyrightyear{2020}
\acmYear{2020}
\setcopyright{acmlicensed}
\acmConference[ICPE '20]{Proceedings of the 2020 ACM/SPEC International Conference on Performance Engineering}{April 20--24, 2020}{Edmonton, AB, Canada}
\acmBooktitle{Proceedings of the 2020 ACM/SPEC International Conference on Performance Engineering (ICPE '20), April 20--24, 2020, Edmonton, AB, Canada}
\acmPrice{15.00}
\acmDOI{10.1145/3358960.3379136}
\acmISBN{978-1-4503-6991-6/20/04}

\begin{document}

\fancyhead{}

\title{GAPP: A Fast Profiler for Detecting Serialization Bottlenecks in Parallel Linux Applications}

\author{Reena Nair}
\authornote{Both authors contributed equally to this research.}
\email{r.nair16@imperial.ac.uk}
\author{Tony Field}
\authornotemark[1]
\email{t.field@imperial.ac.uk}
\affiliation{%
  \institution{Imperial College London}
  \streetaddress{South Kensington}
  \city{London}
  \postcode{SW7 2AZ}
}

%
\renewcommand{\shortauthors}{Trovato and Tobin, et al.}

%
\begin{abstract}
We present a parallel profiling tool, \toolname{}, that identifies
serialization bottlenecks in parallel Linux applications arising from load
imbalance or contention for shared resources . It works by tracing kernel
context switch events using kernel probes managed by the extended Berkeley
Packet Filter (eBPF) framework.  The overhead is thus
extremely low (an average
4\% runtime overhead for the applications explored), the tool requires no program
instrumentation and works for a variety of serialization bottlenecks.  We evaluate \toolname{} using the \emph{Parsec3.0}
benchmark suite and two large open-source projects: \emph{MySQL} and
\emph{Nektar++} (a spectral/hp element framework). We show that \toolname{} is able to reveal a wide range
of bottleneck-related performance issues, for example arising from
synchronization primitives, busy-wait loops, memory operations, thread
imbalance and resource contention.
\end{abstract}

%
%

%

%

%
\maketitle
\section{Introduction}
A key challenge for multi-core program developers is
identifying and fixing performance problems and this is exacerbated in
multi-core systems because of competition for shared resources. Such resources
may be physical, e.g. a hardware accelerator which can only be accessed by one
thread of computation at a time, or logical, e.g. a lock protecting access to
parts of, or the entirety of, a shared data structure. The speed-up achievable
in such cases is thus limited by the inherent serialization that occurs, as
governed by Amdahl's law.

The resource(s) that experience the greatest contention are called the {\em
bottleneck resources}, or just {\em bottlenecks}, akin to the terminology used
in queuing theory. Such resources have a queue of service requests
associated
with them and the immediate symptom of a bottleneck is excessive queuing at
one
or more of those resources. The key insight from queuing theory is that performance can
{\em only} be improved significantly by ``fixing the bottleneck'', i.e. by
reducing the load placed on the resource, e.g.  by accessing it less often or
holding it for less time, on average. Moreover, when there are multiple bottlenecks in a system, we can rank them
based on a metric that takes into account how long the bottleneck is active and the length of the queue. 

This paper presents a new bottleneck detection tool, called \toolname{}
(Generic Automatic Parallel Profiler), which automatically pinpoints
the line(s) of code that represent bottlenecks in a parallel Linux
application. In contrast to some other bottleneck detection systems
which work by instrumenting specific languages, libraries or
synchronization primitives~\cite{yoga2017fast, yu2016syncprof, ul2017syncperf, Tallent:2010:ALC:1693453.1693489, david2014continuously}, 
\toolname{} uses lightweight instrumentation at the kernel level. Furthermore, all bottleneck analysis in \toolname{} is performed at run time, which avoids the need to generate, and
subsequently process, potentially expensive trace files. Indeed, a key
objective in \toolname{}'s design has been to minimize overhead, both during
program execution and in post processing.

The idea is to use the extended Berkeley Packet Filter (eBPF)
framework~\cite{corbet_2014}~(eBPF has been a standard feature of Linux since v4.1) to trace context-switching events inside the Linux
kernel
and keep track of the number of active threads at all times during a program's
execution. We then identify the bottleneck thread(s) by taking account of both
the duration of the thread's time-slices and the degree of parallelism
exhibited
whilst each is executing. We use a weighting algorithm similar to that
described
in~\cite{joao2012bottleneck, du2013criticality} and later also
in~\cite{kambadur2014parashares} in order to do this. The information captured
at context switching events is augmented by information from a lightweight
sampling-based profiler that identifies the program counter
location(s) that correspond to the bottleneck in the code. This is also
implemented using \emph{eBPF}. If the program under
test is compiled to allow stack tracing, we are then able to use the
information
gathered to pinpoint the line(s) of source code that constitute the
bottleneck and which should therefore be the target for optimization.

We make the following contributions:
\begin{itemize}
\item We present \toolname,
a tool to identify arbitrary serialization bottlenecks in parallel Linux
applications based on a weighted {\em criticality} metric (\cm)
(Section~\ref{sec:overview}).  The tool has very low overhead, requires no instrumentation
and works for a variety of bottlenecks, where there is reduced parallelism.

\item We show how
eBPF can be used to build a lightweight sample-based profiler that, when used
in
conjunction with the bottleneck detector, is able to point the programmer at
the
line(s) of source code where the bottleneck is located
(Section~\ref{sec:sampling-probe}).
\item We evaluate the effectiveness of
\toolname{} using \emph{MySQL} and \emph{Nektar++}~\cite{cantwell2015nektar++}
and benchmarks from the Parsec 3.0 suite
(Section~\ref{sec:evaluation}). We confirm known bottlenecks in Parsec 3.0
and also expose and discuss new, previously unreported
bottlenecks.

Performance experiments show
that \toolname{} introduces an average overhead of circa 4\% (maximum circa 13\%)
for the applications considered.
\end{itemize}

\section{Thread Criticality} 
\label{sec:overview}

Existing parallel profilers identify bottlenecks by ranking code regions in
terms of their share of parallelism~\cite{kambadur2014parashares}, optimization
opportunity~\cite{Curtsinger2015} and asymptotic parallelism, i.e the ratio of
total work to critical work~\cite{yoga2017fast}. \toolname{} identifies
bottlenecks by ranking thread execution slices in terms of their execution
time,
weighted by the number of active threads. We use the term \emph{Criticality
Metric (CMetric)} for this weighted metric, borrowing the terminology
from~\cite{du2013criticality}. The advantage of this metric is that it
distinguishes threads that execute for short periods with low parallelism from
those that execute for longer periods with a similar low degree of parallelism.

\subsection{Calculating the \emph{CMetric}} 

\begin{figure}
    \includegraphics[width=0.6\columnwidth,height=3.2cm]{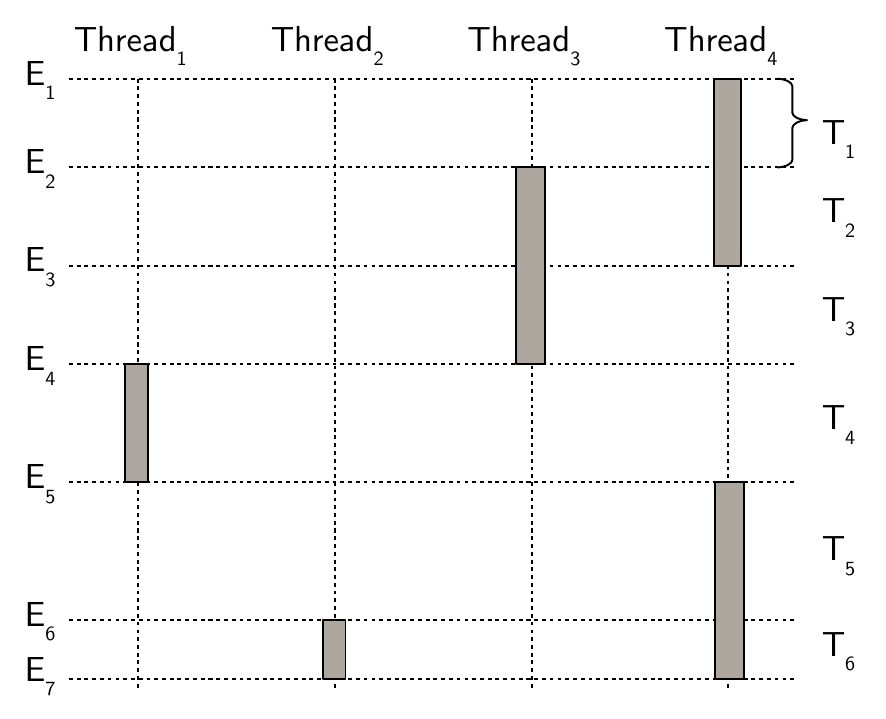}
\caption{Switching intervals}  
\vspace{-5 mm}
\label{fig:CMetric}
\end{figure}

The \cm{} is calculated at the end of each execution timeslice. Timeslices are further divided into \emph{switching
intervals} demarcated by instants when {\em any} thread in the application
changes state from the active to inactive state, or vice versa, as it is only
this action that can change the degree of parallelism (threads are considered to be active in TASK\_RUNNING state and inactive otherwise). Figure~\ref{fig:CMetric}
shows an example trace of a multithreaded application with four threads. The
$E_i, 1 \leq i \leq 7$ are context switching events and the $T_i, 1 \leq i \leq
6$ are the time intervals between consecutive events, $E_i$ and $E_{i+1}$.
Thus,
for example, \emph{Thread$_3$'s} timeslice spans two switching intervals: $T_2$
and $T_3$.

The \cm{} for every active thread is updated at the end of each switching
interval. For example, in the interval $T_2$ in Figure~\ref{fig:CMetric}, there
are two active threads, and hence the \cm{} for that interval is
$\frac{T_2}{2}$, which is added to the individual \cm{} of both \emph{Thread$_3$} and
\emph{Thread$_4$}. The \cm{} for a timeslice of execution is the sum of the
contributions from each switching interval that occurred during the timeslice.

As an example, if a thread's timeslice spans the switching intervals $j, j+1,
...,
k$ then the \cm{} at the end of its timeslice will be
$\sum_{i=j}^k\frac{T_i}{n_i}$, where $n_i$ is the number of application threads
that are active during interval $i$.

In~\cite{du2013criticality} the idea was to use special-purpose hardware to
keep
track of the criticality metric with low cost. Here, we seek to achieve a
similar effect in software, by exploiting existing mechanisms for tracing, and
subsequently filtering, context-switch events in the Linux kernel.

\section{Application tracing using eBPF}

Calculating the \cm{} requires tracking the duration and the number of active threads for each switching interval. In order to avoid instrumenting thread library primitives, and to capture scheduling activities triggered by other events, for example I/O, we keep track of active threads and their execution
duration by tracing context-switch events using the eBPF framework,
which provides a fast and secure mechanism to selectively trace kernel
events~\cite{fleming_2017, sharma2016enhanced}.

Kernel events are traced by eBPF using place holders in the kernel, called {\em
tracepoints} which, when enabled, can be monitored by attaching user defined
probe functions to them ~\cite{Desnoyers:2009:Online}. \toolname{} attaches a
probe function to such a tracepoint, \verb|sched_switch|, which is triggered at
each context switch. This probe function keeps track of the number
of active threads and also calculates the \cm{} for each switching interval, as
specified above.

Information about context-switches are maintained and shared with the user-space using eBPF \emph{maps}. \emph{Maps} can be global~(shared between all cores) or per-cpu~(local to a core). The \toolname{} architecture is shown in Figure~\ref{fig:GAPP-arch}.

In order to compute the \cm{} for the application threads, {\toolname}'s probe
functions set up the following eBPF maps:

\begin{table}[h]
{\notsotiny{}
\centering
\begin{tabular}{|l|l|}
\hline
Map name & Description\\
\hline
\verb|cm_hash| & Global hash map to store the \cm{} of each thread\\[3pt]
\verb|global_cm|& Global scalar - 
cumulative sum of {\cm}s across all switching events\\[3pt]
\verb|local_cm| & Local scalar - records value of \verb|global_cm| when a thread switches in\\[3pt]
\verb|thread_count|& Global scalar - keeps track of the no. of active threads at any time\\[3pt]
\verb|total_count| & Global scalar - stores the total number of threads in the application\\[3pt]
\verb|thread_list| & Global hash - for each thread, $0$ if thread is inactive and $1$ if active\\[3pt]
\verb|t_switch| & Local scalar - stores the timestamp of the most recent switching event\\[3pt]
\hline
\end{tabular}
\caption{eBPF Maps for calculating CMetric}
}
\vspace{-5mm}
\end{table}

\subsection{Identifying application threads}
The \verb|sched_switch|
probe function need to be executed only if either or both of the threads being
switched in/out belong to the application. We therefore capture and store to
\verb|thread_list|, the identifiers of the application tasks, by attaching
additional probes to \verb|task_rename| and \verb|task_newtask| tracepoints, which are invoked when new processes/threads are created. They also increment \verb|total_count| as new tasks are created,
and a probe attached to \verb|sched_process_exit| tracepoint decrements it,
when
threads exit. Hence at any time, \verb|total_count| represents the total number
of threads in the application.

\subsection{Maintaining the number of active threads}
The \verb|thread_count| is maintained in part by the probe 
function attached to the \verb|sched_switch|
tracepoint. The arguments to this tracepoint include a. \verb|prev_pid|: id of the thread being switched out and b. \verb|next_pid|: id of the thread being switched in.

\noindent The \verb|thread_count| is incremented whenever the \verb|next_pid|
belongs to the application and was marked inactive in the \verb|thread_list|.
Threads that were already in the running state do not alter the
\verb|thread_count| when they are switched in. \verb|Thread_count| is
decremented only when \verb|prev_pid| is switched out to an inactive state.

When a waiting thread is woken up
it will eventually be switched in, but there may be a delay between the wake-up
and the context switch. During this time the thread is runnable, so it should be marked active in \verb|thread_list| as soon as it is woken up.  We therefore also trace wake-up events by attaching a
probe function to the \verb|sched_wakeup| tracepoint. This decrements \verb|thread_count| if the thread being woken up belongs
to the application.

\section{Bottleneck Detection}
\label{sec:CMetric Calculation}
\subsection{Calculating the CMetric}

Bottlenecks are identified by ranking execution timeslices in terms of their
\cm, as detailed in Section~\ref{sec:user-probe} below.  At each context switch
event, \verb|global_cm| is updated from the kernel probe thus: \verb|global_cm += (t-t_switch)/thread_count|, where \texttt{t} is
the current time and \texttt{t\_switch} is the timestamp of the last switching
event.  Notice that \verb|t-t_switch| corresponds to the length of the latest
switching interval ($T_i$ for some $i$ in Figure~\ref{fig:CMetric}).  When an
application thread is switched out, the \verb|cm_hash| entry for that thread
(identifier \verb|prev_pid|) is then updated thus:
\begin{verbatim}cm_hash[prev_pid] += global_cm -local_cm\end{verbatim}
Note that if  $n_i$ is the number of active application threads in the $i^{th}$ interval and  the
current timeslice spans the switching intervals between $j$ and $k$ inclusive,
then the right-hand side above is equivalent to $\sum_{i=j}^k\frac{T_i}{n_i}$,
as required.

If the thread being switched in belongs to the application then we store the value of \verb|global_cm| to prepare for the next update to
\verb|cm_hash|, viz.  \verb|local_cm = global_cm|.

\subsection{Stack traces}
At the end of a time-slice, we need to determine if it represents a potential bottleneck, and if so, we wish to pinpoint the line(s) of code
that correspond to the bottleneck and also the call path that led to it
by means of stack traces.  The latter is important as it may be that a
bottleneck appears only for a specific call path, even though other paths may
lead to the same line of code.

A stack trace is recorded if the average level of
parallelism during the timeslice is below a given target threshold, $
N_{min}$,
a tunable parameter, thus indicating
the execution of a potential bottleneck at some stage during that timeslice. To
do this we compute the weighted average of the number of active threads during a
timeslice, \verb|threads_av|, similar to the calculation of the \cm{}
detailed
above.  A stack trace is triggered if \verb|threads_av| $< N_{min}$.

\begin{figure}
    \includegraphics[width=0.7\columnwidth, keepaspectratio=true]{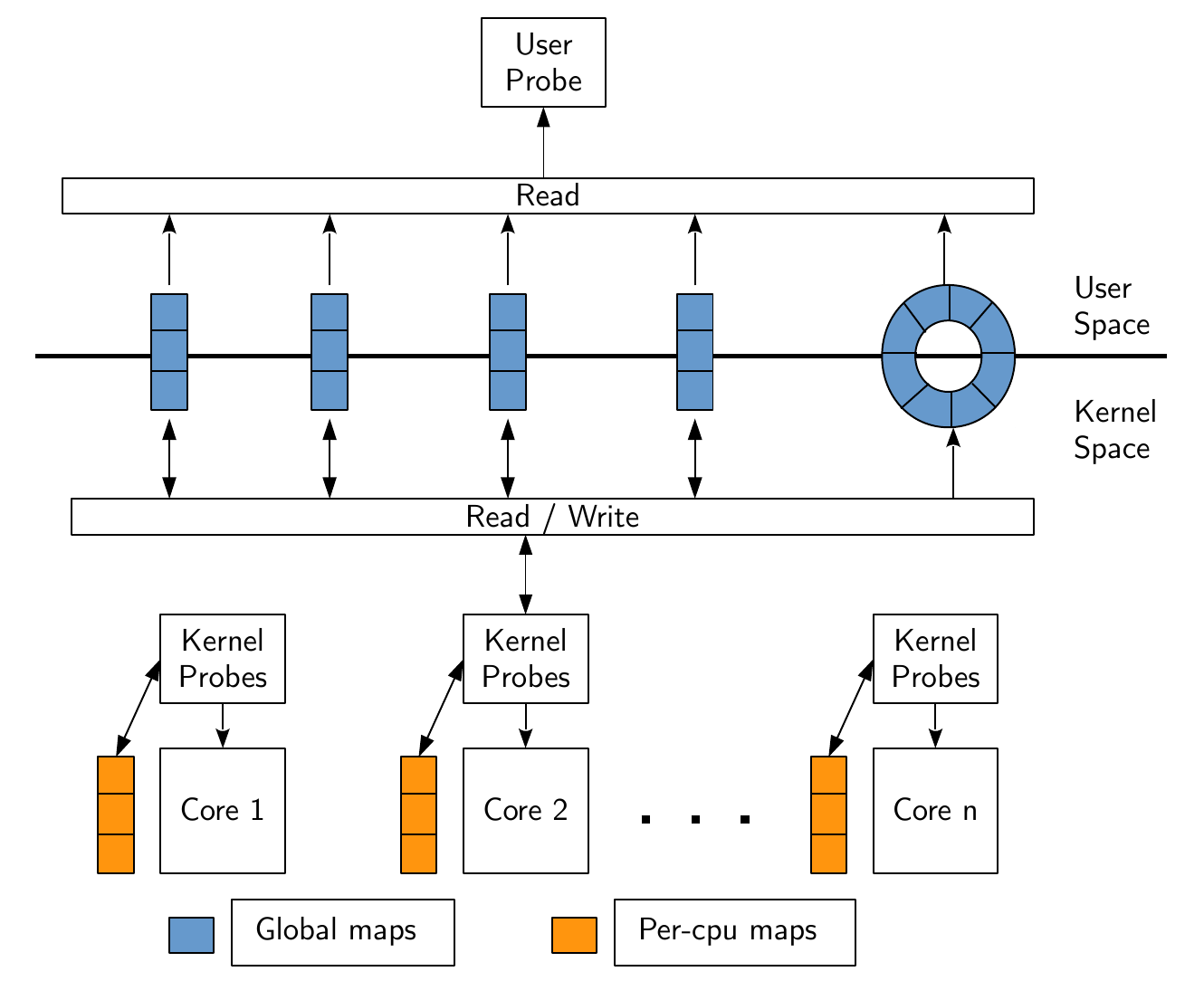}
\caption{GAPP Architecture}  
\vspace{-5 mm}
\label{fig:GAPP-arch}
\end{figure}

At the end of a time-slice, if a stack trace has been triggered, then the thread
id, \cm{} and stack trace are written to a circular buffer managed by
\emph{eBPF}~(Figure~\ref{fig:GAPP-arch}). This buffer is readable from a user-space probe that runs in
parallel with the application threads.  To
avoid having to store every entry in deep stack traces we only record the top
$M$ (user-specified) entries in each.

\subsection{Sampling probes} \label{sec:sampling-probe} The above scheme works
well if threads happen to be executing bottleneck code when a context switch
happens. However, in many cases bottleneck code can be expected to
complete its execution before the next
context-switch event, i.e. before the next stack tracing opportunity. Because of this we have found that stack traces taken at context
switches alone are often not enough to pinpoint the root cause of a bottleneck.

To address this problem, we use \emph{eBPF} to add an additional lightweight
periodic sampling probe, with period $\Delta t$ (another tunable parameter) which
additionally records the instruction pointer whenever the absolute number of
active threads, as given by \verb|thread| \verb|_count|, is less than $N_{min}$. 
If these conditions are met then the thread id and instruction pointer are written
to the same \emph{eBPF} circular buffer referred to above.

\subsection{User space probe}
\label{sec:user-probe}
 The user-space probe,
written using the \emph{BPF Compiler Collection} (bcc)~\cite{fleming_2017} tool
kit, runs in parallel with the application threads and communicates with the
kernel probes via the circular buffer.

Instruction pointers from the sampling probe are read and assembled in a hash
map that is indexed by the thread id. When a \cm{} and stack trace associated
with the same thread id are read from the buffer the information accumulated is
used to populate three locally-managed hash maps, indexed by a unique id,
\verb|ts_id|, associated with the timeslice.  These hash maps contain the \cm,
the call path corresponding to the stack trace, and a list of addresses which
include those from the sampling probe. These addresses are the candidates for
bottleneck lines of code. If a stack trace is {\em not} triggered at the end of
a timeslice then a special entry is written to the circular buffer, along with
the current thread id, which instructs the user probe to reject any instruction
pointers from sample probes associated with that thread.

When the program terminates, the user probe enters a post-processing phase
which
seeks to determine the instruction pointers that have the highest collective
\cm. At this point there will, in general, be a number of identical call paths
whose \verb|ts_id| indices are included in the list of \cm{} entries. If \verb|i|
and \verb|j| are two such indices, then the corresponding entries in the
address
lists  and \cm{} map are {\em merged} by: \textit{a}.  summing the \cm{} values
at
indices \verb|i| and \verb|j|, \textit{b}.  combining the addresses at indices
\verb|i| and \verb|j| to generate a frequency table for those addresses.

At the end of the merge process, we are thus left with two hash maps, each
indexed by a unique call path; one with the accumulated \cm{} for
the call path and the other with the list of sampled addresses for the same
call path.  The entries with the top $N$ total \cm{}s are then taken as the
bottlenecks.

The reason why the top $N$ entries are chosen instead of the top one
alone is because one call path could be a subset of another, as would occur if
a
context-switch happened during execution of one function that indirectly calls
another that contains a bottleneck.

Finally, the items in the address map(s) are mapped to function names and the
lines of code associated with them by calling the Linux \verb|addr2line|
utility. The final profile is presented as a frequency table of functions and
lines of code as illustrated in Figure~\ref{fig:MySQL}.

\subsub{Critical timeslices with no samples}
\label{sec:no-samples}
Sometimes, a timeslice may be identified as critical, but the sampler may fail
to gather the instruction pointer(s) that define the location of the
bottleneck.
This can happen due to two reasons: either the timeslice was too short, so that
the sampler missed it, or the sample belonged to a shared library or kernel
code
and hence could not be mapped to the source code. In the absence of any
samples,
the next best thing is to add the address at top of the stack, i.e., the return
address of the caller, to the list of samples. In summary, the top stack
address
is attached to the samples, if \textit{a}. the sample count is zero and
\textit{b}. the active thread count is less than or equal to $N_{min}$, when
the
thread is switched out. Such samples are labelled as being from stack top to
help the user interpret results correctly.

\section{Evaluation}\label{sec:evaluation}
We now detail a series of experiments that aim to achieve: a) Evaluate \toolname's ability to pinpoint bottlenecks in parallel applications and b) Quantify the overhead of the tool for various benchmarks.

\subsection{Experimental Setup}
 We evaluate \toolname{} using applications from
the Parsec 3.0 benchmark suite~\cite{bienia2008parsec} and two larger
open-source projects: \emph{MySQL} and \emph{Nektar++}\cite{cantwell2015nektar++}, a spectral/hp element framework for solving partial differential equations.  Experiments were performed on a server machine with four AMD Opteron 6282SE eight-core/16-thread CPUs
(with
a total capability to run 64 threads in parallel) and 128GB of RAM.  The system
runs on Linux 4.15 with the \verb|bcc| toolkit installed. The applications were
compiled with \verb|gcc| version 7.3.0 at the \verb|-O3| optimization level.
The
compiler options \verb|-g| and \verb|-fno-omit-frame-pointer| were enabled to
aid effective call stack retrieval. Recent versions of \verb|gcc| generates
position-independent executable by default. To map addresses to source code
using \emph{addr2line}, this behaviour need to be overridden through compiler
and linker options \verb|-fno-pie| and \verb|-no-pie| respectively.

In the initial set of experiments we used $N_{min} = n/2$ where $n$ is the
number of application threads, and a sampling period of $\Delta t = 3ms$. The sensitivity of the framework to these parameters is evaluated in
the \verb|README| file of \toolname{}'s Github repository~\cite{GAPP}.

\subsection{Results and Analysis}

In this section, we analyze the results of profiling different parallel applications with \toolname{}.

\subsection*{Parsec 3.0 benchmark}

We evaluated \toolname{} with 11 multi-threaded applications from the Parsec3.0~\cite{bienia2008parsec} benchmark suite. 
The applications were executed with 64 threads on the \verb|native| input set, as we were using a 64-core machine, although the framework does not restrict the number of threads in any way. All the applications except \emph{Freqmine}
were compiled with the \emph{Pthreads} library. \emph{Freqmine} uses the \emph{OpenMP} threading library. 

Out of the 11 applications used for evaluation, two, namely \emph{Dedup} and \emph{Ferret}, are task-parallel; the remaining are data-parallel. \toolname{} is able to identify not only previously established bottlenecks, but also several that have not previously been reported.
We compare the results with those from previous studies and show that
\toolname{} is able to detect, and pinpoint, serialization bottlenecks
resulting from, e.g. synchronization primitives, workload imbalance, busy-wait loops and contention. 

Since these applications have been extensively analyzed in previous studies, we
only report results which add to the existing analysis. 
The results obtained for other applications are listed in 
Table-\ref{tab:Results}, which summarizes the critical functions along with original program execution time (T), the \toolname{}
overhead as a percentage of
the original time (O/H), the proportion of critical timeslices to the total timeslices as a percentage (CR), the memory usage of the tool (M) in MB and the post processing time (PPT) in seconds.

\begin{table*}
{\small
\centering
\begin{tabular}{|l|l|l|l|p{2cm}|p{1cm}|p{1cm}|}
\hline
Application & Critical Functions identified by \toolname & O/H & T (s) & CR & M (MB) & PPT (s)\\
\hline

Blackscholes & CNDF() &  < 1\%  & 29.4 & 470 (2\%) & 109 & 0.02\\

Bodytrack & OutBMP, RecvCmd & 5\% & 21.3 & 6823 (0.5\%) & 112 & 0.4\\

Canneal & netlist\_elem::swap\_cost & 2.2 \% & 62 & 267 (0.06\%) & 112 & 0.1\\

Dedup & deflate\_slow & 12\% & 13 & 362544 (40\%) & 372 & 3.3\\

Facesim & Update\_Position\_Based\_State\_Helper & 4.4\% & 59 & 334 (0.004\%) & 118 & 0.8\\

Ferret & dist\_L2\_float & 3\% & 30 & 42127 (51\%) & 132 & 1.2 \\

Fluidanimate &  parsec\_barrier\_wait & 2\% & 37 & 11512 (1\%) & 112 & 0.3\\

Freqmine & FPArray\_scan2\_DB & 2\% & 34 & 11721 (13\%) & 110 & 0.6\\

Streamcluster & parsec\_barrier\_wait, dist  & 5.6\% & 201 & 2246172 (10.6\%) & 784 & 2.5\\

Swaptions & HJM\_SimPath\_Forward\_Blocking & 1\% & 8.6 & 43 (0.07\%) & 111 & 0.2 \\

Vips & imb\_LabQ2Lab & 4\% & 14 & 10460 (3.2\%) & 118 & 2.3\\

MySQL & fil\_flush, sync\_array\_reserve\_cell & < 1\% & 60 & 825 (0\%) & 194 & 3\\

Nektar++ & dgemv\_ & 9\% & 37 & 189990 (16\%) & 446 & 2.1\\
\hline
\end{tabular}
\caption{Critical functions identified with overhead (O/H), execution time (T),
Number of critical time slices with proportion of Critical timeslices to total
timeslice (CR), memory usage in MB (M) and post-processing time (PPT).}
\label{tab:Results}
}
\end{table*}
\subsub{Bodytrack}
\verb|Bodytrack| is a vision application that tracks the 3D pose of a
marker-less human body with multiple cameras through an image sequence. The
application uses worker threads to process video frames as per the
commands from the parent thread. \toolname{} identified \verb|OutputBMP()| and \verb|RecvCmd()| as the top critical functions. In \verb|RecvCmd()|, the worker threads wait on a conditional variable till they receive commands from the parent thread.

\begin{figure}[ht]
\centering
\includegraphics[width=0.8\columnwidth, height=3.5cm]{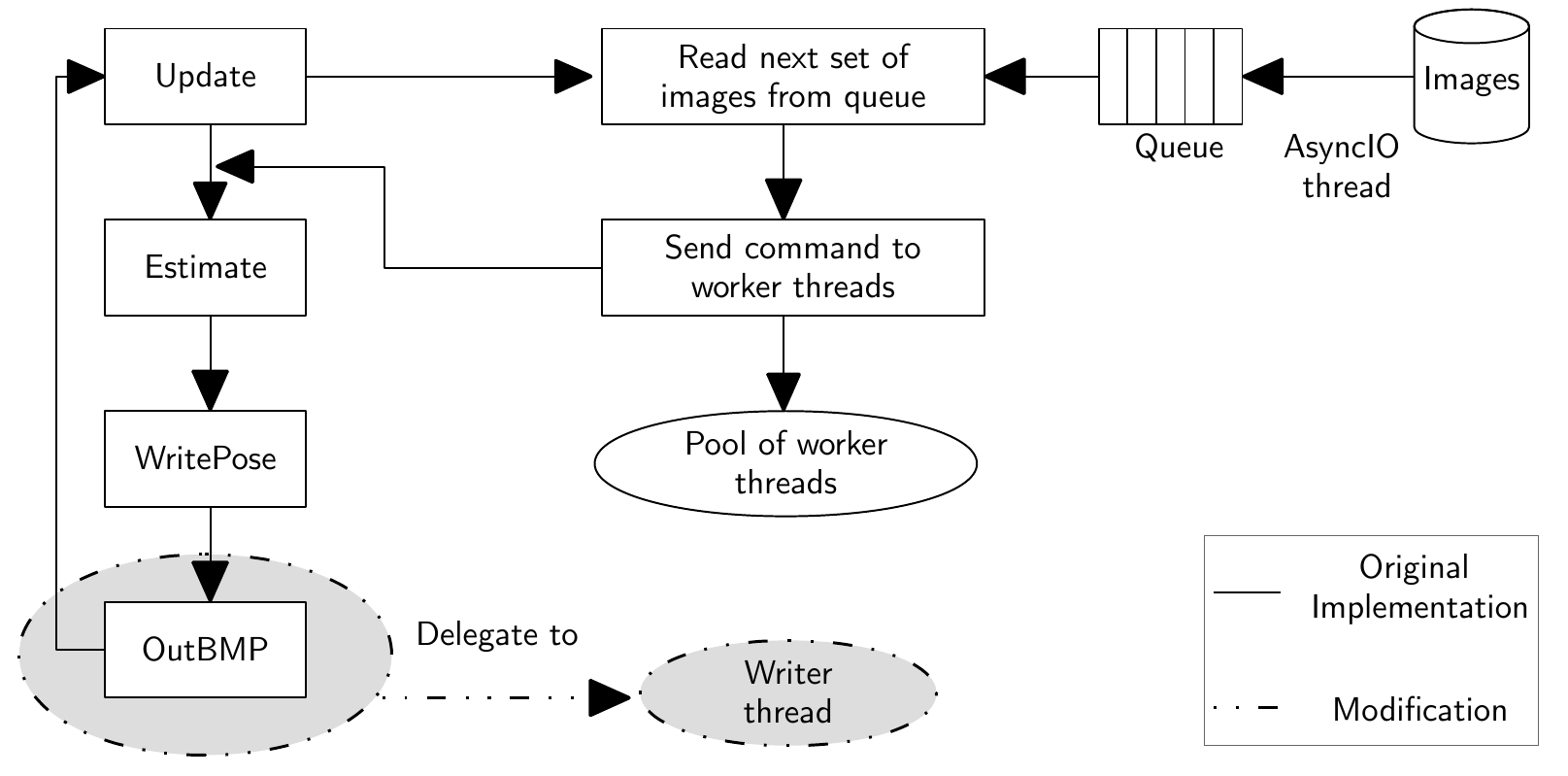}
\caption{General logic of \emph{Bodytrack}}
\label{fig:bodytrack_logic}
\end{figure}

The \verb|OutputBMP| function invoked by the parent thread, generates the camera image and saves it in BMP format. While the images are processed by \verb|OutputBMP|, the worker threads are waiting on the conditional variable in the \verb|RecvCmd| function.  We commented out \verb|OutputBMP| function and profiled all threads with \toolname{}, which showed 45\% reduction in the number of samples from \verb|RecvCmd|. This confirms that the \verb|OutputBMP| function is indeed the bottleneck.

We offloaded the \verb|OutputBMP| function from the parent thread to a new
thread called \emph{writerThread}. Consequently, the parent passed commands to
the worker threads at a faster pace, thereby reducing the worker's waiting
time. The modification improved the performance of \emph{Bodytrack} by 22\%.
This bottleneck has not been reported before. Figure~\ref{fig:bodytrack_logic} outlines the logic of \emph{Bodytrack}, before and after optimization.

\subsub{Ferret and Dedup}
\label{sec:dedup}
\emph{Ferret} and \emph{Dedup} are task-parallel applications designed with different pipeline stages,
with a queue connecting each stage to the next one.

\emph{Ferret} implements content-based similarity search on images, audio, 3D shapes etc, and is implemented with six pipeline stages. The first
and last phases, which perform I/O, are serial phases. The middle four stages
implement query image segmentation, feature extraction, indexing and ranking
respectively. We executed \emph{Ferret} with $15$ threads in each of the
parallel phase making a total of $62$ threads for the whole execution. The top
critical functions identified by \toolname{} were invoked by \verb|emd()| 
(Earth Mover's Distance) which finds the pair-wise distance between query
image and candidate images, and also forms the core of the \emph{ranking} phase. 

\begin{figure}[ht]
\begin{minipage}[b]{\columnwidth}
\centering
\includegraphics[width=0.95\columnwidth, height=4cm]{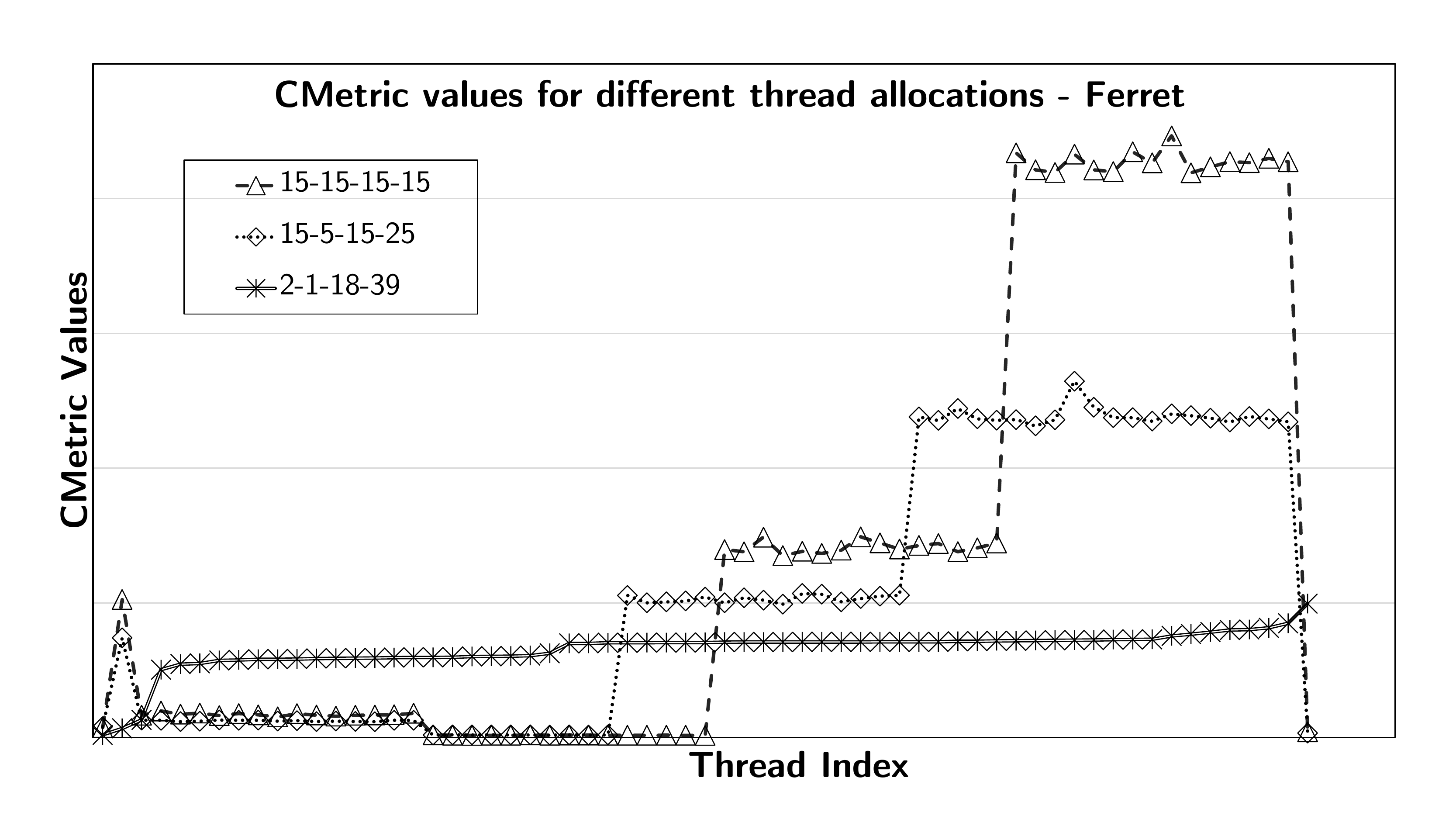}
\end{minipage}
\caption{CMetric for different thread allocations - Ferret}
\vspace{-2mm}
\label{fig:Thread Reallocation-Ferret}
\end{figure}

As evident from Figure~\ref{fig:Thread
Reallocation-Ferret}, \verb|Ferret| exhibited huge imbalance in
\cm{} among the threads(We have used a line chart to represent this discrete data
in order to highlight the variation in \cm among the phases, for a particular
thread allocation). The threads with higher \emph{CMetric} belonged to the
\verb|ranking| phase as was evident from the critical functions identified by
\toolname{}~(see Table~\ref{tab:Results}:Ferret).

We redistributed the load among the parallel phases in \emph{Ferret}, until we obtained a uniform \cm{} for the threads, as shown in figure~\ref{fig:Thread Reallocation-Ferret}. Threads were reallocated as 2-1-18-39 among the parallel phases, which improved the run time of \emph{Ferret} by 50\%, almost double the speed-up achieved by the reallocation of 20-1-22-21 suggested by \cite{Curtsinger2015}~(23\% speed up on the test bed).
 
\emph{Dedup} is designed with five pipeline stages,
viz., \emph{Fragment},  \emph{FragmentRefine}, \emph{Deduplicate},
\emph{Compress} and \emph{Reorder}. The first and last stages perform I/O with
a single thread, and the rest parallelize the task among a pool of worker threads.
In our experiments, \emph{Dedup} was configured to run with $20$ threads in the
intermediate stages to make a total of $62$ threads for the application.

With an initial thread allocation of 1-20-20-20-1, the \verb|write_file()| function from the \emph{Reorder} phase, and the \verb|deflate_slow| function from the \verb|Compress| phase were identified as the top critical paths. The sequential phase, \emph{Reorder}, which writes the compressed chunks to a file, is known to be a bottleneck~\cite{DeSensi:2017:BPP:3154814.3132710}. To accelerate the \emph{Compress} phase, we moved threads from the \emph{FragmentRefine} and \emph{Deduplicate} stages to the \emph{Compress} stage, and ran the experiment with 1-16-16-28-1 threads in the respective stages. However, increasing parallelism in the \emph{Compress} stage increased the run time, which indicates possible contention in the particular stage. We decreased the number of threads in the \emph{Compress} stage from $20$ to $15$ (thread allocation 1-20-20-15-1), and this improved the run time by 14\%.

The workload imbalance among threads in \emph{Dedup} and \emph{Ferret} was
reported in~\cite{ul2017syncperf} for a 16 core machine, which also suggests an optimal thread allocation for the two applications. We have included it here to show how \toolname{} can be used to tune the load among threads in such cases.

\subsection{GAPP on real world applications}

We have also tested \toolname{} with two large applications viz.,
\emph{Nektar++} and \emph{MySQL}.
\emph{Nektar++} is a multi-process application that uses
\verb|OpenMPI|, while \emph{MySQL} is a multi-threaded database management system.

\subsub{Nektar++}
We evaluate \toolname{} on an MPI application, \verb|Nektar++|, which
implements scalable PDE solvers using the spectral/hp element
method~\cite{cantwell2015nektar++}. We focused the evaluation on the
\texttt{Incompressible Navier-Stokes Solver}~(IncNSS), which was configured to run with $16$ processes on a cylindrical surface. The solver divides the complex problem into several small elements represented as an unstructured mesh, partitions the mesh and assigns each partition to an individual process.

Message passing in \verb|Nektar++| is implemented with \verb|OpenMPI|, and by
default, MPI processes are configured to run in ``aggressive'' mode. In this
mode, rather than blocking waiting for messages, processes spin on a busy-wait
loop. Hence the initial profile generated by \toolname{} exhibited uniform
\cm{} for all processes, as they are never idle. However, this behaviour masks any load
imbalances in the application.
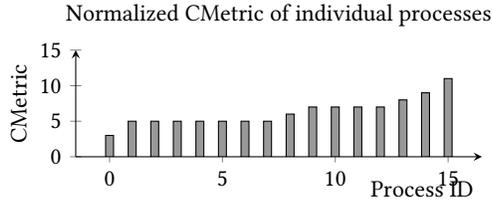
\begin{figure}
\begin{minipage}[c]{0.5\columnwidth}
\centering
\begin{tikzpicture}[trim left=0.75cm]
%
\begin{axis}[
    ybar,
	ymin = 0,
	ymax = 15,
    height=3 cm,
    width = 0.5\columnwidth,
	axis lines=center,
	axis lines* = left,
	x=0.3cm, 
    enlarge x limits=0.10,
	bar width=3pt,	
 	ylabel = {CMetric},
 	y label style={at={(axis description cs:-0.1,.5)},rotate=90,anchor=south},
	xlabel = {Process ID},
    every axis x label/.style={at={(1,-0.15)},anchor=north east},
    title = {Normalized CMetric of individual processes},
	]
	]
\addplot [fill=black!40!white] table [y=CMetric,  x expr=\coordindex] {NektarCM.dat};
\end{axis}
\end{tikzpicture}
\end{minipage}
\vspace{-5mm}
\caption{CMetric of individual processes - Nektar++}
\vspace{-3mm}
\label{fig:Nektar_CM}
\end{figure}
To disable busy waiting, we recompiled \verb|Nektar++| with \verb|MPICH|, an
alternate implementation of the MPI standard, enabling the
\verb|--with-device=ch3:sock| option. This revealed a substantial
non-uniformity in load, as shown in Figure~\ref{fig:Nektar_CM}.  

A likely cause of such load imbalances is non-uniform partitioning. To test
this we artificially created a structured mesh for a cuboid surface and uniformly
partitioned it among eight processes. The use of a structured mesh makes it easier
to generate uniform partitions.
With this partitioning, processes exhibited negligible variation in \cm{},
which confirms that the imbalance in the cylidrical solver was indeed non-uniform
partitioning. Fixing this would involve re-engineering the mesh partitioner,
which is beyond the scope of the current work.

\begin{figure}
\begin{minipage}[c]{\columnwidth}
\centering
\begin{tikzpicture}

\begin{axis}[
	title = {Original},
    compat=newest,
    ybar,   
    ymin=0,         
    ymax=100,
    xtick=data, 
    height=0.4\textwidth,
    axis lines=center,
    xlabel = \empty,
    x tick label style={rotate=45, anchor=east, align=right},
    ylabel = {Normalized Count},
    y label style={at={(axis description cs:-0.35,.5)},rotate=90,anchor=south},
     title style={at={(0.7,1)},anchor=north},
    xticklabels from table={NektarO3.dat}{Function},
    axis lines* = left, 
    every axis x label/.style={at={(current axis.right of origin)},anchor=north west},
    tick label style={font=\notsotiny},
    ybar=5pt,
    bar width=8pt,
    enlarge x limits=0.3,
    x=0.45cm,
    width=\textwidth,
]
\addplot [fill=black!50!white] table [y=Count, x expr=\coordindex] {NektarO3.dat};
\end{axis}

\begin{axis}[
	at={(0.5\linewidth,0)},
    title = {Optimized},
    compat=newest,
    ybar,   
    ymin=0,         
    ymax=100,
    xtick=data, 
    height=0.4\textwidth,
    axis lines=center,
    xlabel = {Critical Function},
    xlabel style={draw,rectangle,align=left,text width=2cm},
    x tick label style={rotate=45, anchor=east, align=right},
    ylabel = \empty,
     title style={at={(0.7,1)},anchor=north},
    xticklabels from table={openBLAS.dat}{Function},
    axis lines* = left, 
    every axis x label/.style={at={(current axis.right of origin)},anchor=north west},
    tick label style={font=\notsotiny},
    ybar=5pt,
    bar width=8pt,
    enlarge x limits=0.3,
    x=0.45cm,
    width=\textwidth,
    clip = false,
]
\addplot [fill=black!50!white] table [y=Count, x expr=\coordindex] {openBLAS.dat};
\end{axis}
\end{tikzpicture}
\end{minipage}
\vspace{-3mm}
\caption{Bottleneck functions in Nektar++}
\vspace{-3mm}
\label{fig:Nektar}
\end{figure}
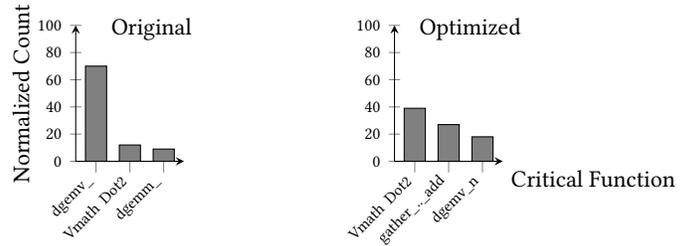

As well as identifying the above load imbalance
\toolname{} pinpointed \verb|dgemv_()|, a matrix multiplication routine exported by the \emph{BLAS} library, as the top critical function in \emph{IncNSS}. We recompiled \emph{Nektar++} with \emph{OpenBLAS}, an
optimized version of the \emph{BLAS} library. This improved the performance of
the application by $27$\% and moved the bottleneck from \verb|dgemv_()| to
\verb|Vmath::Dot2()| function (Figure~\ref{fig:Nektar}). Note that this led to
negligable change in the observed load imbalance.
Note also that \verb|dgemv_()| doesn't represent a serialization bottleneck -- it is simply an
expensive function that happened to be executing with reduced parallelism. 

\begin{figure}
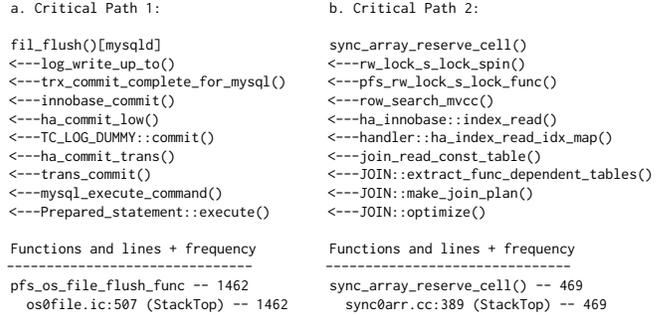

\begin{minipage}{0.5\columnwidth}
\notsotiny
\begin{verbatim}
a. Critical Path 1:

fil_flush()[mysqld]
<---log_write_up_to()
<---trx_commit_complete_for_mysql()
<---innobase_commit()
<---ha_commit_low()
<---TC_LOG_DUMMY::commit()
<---ha_commit_trans()
<---trans_commit()
<---mysql_execute_command()
<---Prepared_statement::execute()

Functions and lines + frequency
-------------------------------
pfs_os_file_flush_func -- 1462
  os0file.ic:507 (StackTop) -- 1462
\end{verbatim}
\end{minipage}%
\begin{minipage}{0.5\columnwidth}
\notsotiny
\begin{verbatim}
b. Critical Path 2:

sync_array_reserve_cell()
<---rw_lock_s_lock_spin()
<---pfs_rw_lock_s_lock_func()
<---row_search_mvcc()
<---ha_innobase::index_read()
<---handler::ha_index_read_idx_map()
<---join_read_const_table()
<---JOIN::extract_func_dependent_tables()
<---JOIN::make_join_plan()
<---JOIN::optimize()

Functions and lines + frequency
-------------------------------
sync_array_reserve_cell() -- 469
  sync0arr.cc:389 (StackTop) -- 469
\end{verbatim}
\end{minipage}
\caption{Critical Paths - MySQL5.7}
\vspace{-3mm}
\label{fig:MySQL}
\end{figure}

\subsub{MySQL}
We profiled \emph{MySQL 5.7} with \toolname{}, while executing the
\verb|OLTP_Read_Write| workload from \emph{Sysbench} benchmark. The top
critical samples were from \verb|pfs_os_file_flush_func()|, which flushes the
write buffers of a given file to disk. The call path shows that this function
was invoked by \emph{InnoDB}, the transactional storage engine for
\emph{MySQL}~(Figure~\ref{fig:MySQL}a.). To optimize \emph{InnoDB} disk I/O, we
increased the buffer pool size to $90$GB (70\% of the total system memory) as
suggested in \cite{mysql-InnoDB}. This improved the transaction rate (measured
in transactions per second) by 19\% and reduced the average latency by 16\%. 

The second most critical samples were from a spin-wait loop in
\emph{sync\_array\_reserve\_cell()} invoked from
\verb|rw_lock_s_lock_spin|, as shown in Figure~\ref{fig:MySQL}b.

Mutexes and read-write locks in MySQL are designed such that the threads poll for them for a short period of time before they block. 
The duration of the spin-wait loop is dependent on a random number between
$0$ and \verb|INNODB_SPIN_WAIT_DELAY|~(a configurable constant). The default
value of this constant is $6$. \cite{mysql-spinLock} advises to set the constant to a higher value to improve performance on machines with large number of cores. We set the constant to a value $30$, and this cumulatively improved the transactions rate by 34\% and reduced the average latency by 25\%. Moreover, the number of cache misses decreased by 10.5\%, when compared with the default delay variable value. This shows better cache performance with increased spin-wait delay. 
Interestingly, optimising the spin-wait delay {\em without} first optimising
the buffer size made negligible difference to the overall performance. This
emphasises the importance of ranking the bottlenecks by criticality and tackling
each in turn.

The bottlenecks detected in \emph{MySQL} were fixed by tuning configuration
parameters, rather than re-factoring the source code.
Existing \emph{MySQL} performance tuning tools are designed to identify similar
optimization opportunities. Nonetheless, the experiments show that \toolname{},
which is a generic tool, is capable of identifying similar architecture dependent bottlenecks.

\subsection{Overhead}
For the 13 applications evaluated, the maximum overhead
incurred was only 13\% (Table-\ref{tab:Results}). 
The overhead induced by a profiler on the application is important, as it can
alter the application's normal behaviour. To minimize the effect of tracing on
program behaviour we collect call paths only at the end of a context-switch and 
only if the time-slice exhibited reduced parallelism; similarly
instruction pointers from the sampling probe.

The overhead was found to be dependent on the ratio of critical time slices to
the total time slices, the depth of the stack traces and the number of distinct
stack traces. This is because \toolname{} caches address-to-symbol mapping, and hence the mapping time will be less when stack traces are identical.

The memory overhead was found to be proportional to the number of samples and the number of critical timeslices/ stacktraces.

\section{Related work}
Bottleneck detection in multithreaded applications
is by no means new and several techniques have been proposed.
Some of them identify parallelism bottlenecks based on critical
sections~\cite{yu2016syncprof,joao2012bottleneck}, critical paths~\cite{chen2012critical}, bottlenecks introduced by specific resource access (synchronization) mechanisms, for example
locks~\cite{Tallent:2010:ALC:1693453.1693489, david2014continuously} or
by native threading libraries, such as POSIX threads~\cite{ul2017syncperf} and
Intel TBB~\cite{yoga2017fast}. These strategies identify execution regions where the resource is acquired and released by instrumenting the source code or binary. The major benefit of using instrumentation mechanisms is the ease to map execution traces to the source code. Information is generated only at instrumentation points thereby limiting the amount of data collected, which makes post-processing faster and easier. However, they can only pin point bottlenecks that arise due to a particular class of resource and are also bound to a particular language or library which restricts their use.

Techniques that work independent of the language or
library commonly rely on stack
traces~\cite{ammons2004finding, han2012performance, Yu:2014:CPR:2541940.2541968}
or core dumps~\cite{altman2010performance} to identify expensive call sequences
or excessive idle time. They are designed to filter and analyze large amounts
of data generated during program execution. 

Yet another category of tools either categorize bottlenecks based on hardware
events~\cite{eyerman2012speedup, 6114195, yoo2012automated} or
limit their scope in identifying the critical thread~\cite{du2013bottle, du2013criticality} alone.

\emph{Bottlegraphs}~\cite{du2013bottle} and \emph{Criticality Stacks}~\cite{du2013criticality} identify the most critical thread by ranking threads based on the execution time and number of active threads. They were proposed as mechanisms to identify and accelerate critical threads for power/ performance optimizations. 

\toolname's logic bears some similarity to those proposed
in~\cite{du2013bottle,du2013criticality, tallent2009effective}
and~\cite{kambadur2014parashares}. \emph{Bottlegraphs}~\cite{du2013bottle} tracks
active threads using kernel modules that intercept futex calls and system calls
that create, destroy and schedule threads. \emph{Criticality
Stacks}~\cite{du2013criticality} proposes a hardware approach to calculate the
\emph{Criticality Metric} and identify the critical threads which are diverted
to faster cores for performance and energy optimization. While both are capable
of identifying critical threads, they do not provide information regarding the
code section or function that causes the bottleneck.
\cite{kambadur2014parashares} ranks all basic blocks in a program based on
their share of parallel execution. Execution information is gathered using
\emph{Parallel Block Vectors}~\cite{kambadur2013parallel}, which uses an LLVM
compiler pass to instrument the \emph{Pthread} library routines in the application. \cite{tallent2009effective} quantifies insufficient parallelism by attaching an \emph{idleness metric} to each call path. It samples a time-based counter, and a signal handler collects the calling contexts, if the application thread is active during the sample.  

While \toolname's logic is similar, it does not require any hardware modification or software instrumentation, produces accurate results as every scheduling event is captured, and achieves the same goal with very low overhead, as calling contexts are gathered only when critical. Moreover, \toolname{} will work with an arbitrary number of threads or when other applications are running concurrently. This is because \toolname{} uses the thread state to determine whether it is active or not, whereas similar approaches proposed in \cite{du2013criticality} and \cite{tallent2009effective} considers a thread to be active only if it is occupying a CPU core. The calculation of degree of parallelism can go wrong in such cases when there are other applications running concurrently or when the number of threads in the application is greater than the number of CPUs. \cite{kambadur2014parashares} calculates the degree of parallelism by instrumenting thread creation/exit/synchronization primitives in the \verb|pthread| library. Hence, if a thread gets blocked for some reason other than synchronization, it will not be captured by the framework.

\toolname{}'s approach is complementary to a recent study on off-cpu
analysis~\cite{zhou2018wperf}, which identifies waiting events critical to
throughput. Information regarding waiting events are recorded by tracking
interrupts and thread switching events using \verb|kprobes|. This information is
used to build a \emph{wait-for-graph}, which is post-processed to determine
threads that are influenced by a waiting event. Even though \toolname{} does not
explicitly state the waiting-relationship among tasks, we have observed that the
same information can be interpreted from stack traces generated during
context-switches. \emph{wPerf} can provide more detailed information than
\toolname{} through its wait-for graphs, but the tradeoff is that it requires substantially longer
post-processing times, e.g. 271.9 sec for \emph{MySQL} as quoted in~\cite{zhou2018wperf}. The reported application
execution time overheads are broadly similar to \toolname{}.

\toolname{}'s results are found to be consistent across multiple runs, and the results can be obtained in a single execution. With tools such as \cite{Curtsinger2015}, we have observed significant variations across experiments on an 8-core machine. This most likely happens because it uses a statistical approach based on sample count to determine the delay inserted in threads and we have found that this can make it difficult to reproduce results from one run to the next. Also \cite{Curtsinger2015} does not report call paths and this can be important when the same code can be invoked from multiple paths. Moreover, none of these tools have been tested on parallel applications that use MPI or similar message passing constructs.

\subsection{Limitations}

Locks or synchronization primitives with low contention may not be identified by \toolname{}. Even though these are not strict serialization bottlenecks, it has been proven that using the proper primitive in such cases can improve the performance~\cite{ul2017syncperf} and hence is a possible optimization opportunity.

\toolname{} may not identify spin locks that spin indefinitely. This is a problem shared by other frameworks that rely on instrumentation, as spin-locks can be implemented using custom-built primitives that instrumentation cannot detect.

As in the case of spin-locks, \toolname{} will not identify bottlenecks in applications that busy wait, such as MPI applications executing in aggressive mode. However, from our experience, disabling the aggressive mode, at least in the development phase, will help identify existing load imbalance among participating processes.

The default behaviour of \verb|gcc|, of generating \emph{position-independent executables}, needs to be overridden for the \verb|addr2line| utility to work properly. This can be overcome by finding a way to map the offset provided by the \verb|sym()| primitive of \verb|bcc| to the source code. 

Also it may be noted that, adding \emph{eBPF} probes to the kernel and removing probes from it require root privileges and hence \toolname{} requires root privileges for its operation.

\section{Conclusion and Future Work}
\label{sec:conclusion}
\toolname{} works ``out of the
box'' in that it can be used to profile an application without the need to
instrument the source code, patch the operating
system or undertake elaborate post-processing of data captured during a
program's execution. Moreover, the verifier in the \emph{eBPF} framework ensures that the probes are safe to attach to a live kernel.

\toolname{} is able to identify 
a range of different
bottlenecks, for example critical sections, execution hot-spots, bottlenecks
identified from
hardware events, busy-wait loops and resource contention. 
\toolname{} has proven to be remarkably good at detecting bottlenecks, but in order
to classify the bottleneck type, e.g. as being due to synchronization or I/O,
more work is required. In order to automate the process of 
bottleneck classification we have recently experimented with tracking
I/O system calls with a view to determining which files, IP addresses etc. an
application interacts with. We have also explored tracing kernel-level
synchronization (``futex'') calls which are used by many higher-level
libraries and macros. For example, by combining GAPP's existing criticality
information with an analysis of futex `wakers' it is relatively easy to 
distinguish critical from non-critical lock holders; this can help to rank
multiple call paths leading to the same lock.
By accessing other hardware counters it
is possible in principle to trace other sources of bottlenecks such as page
faults. 

It would be good to see how \toolname{} behaves with other parallel platforms such as Java, Intel TBB, Cilk etc. While the core concepts in \toolname{} is not attached with any language/library, we need to experiment and see how stack traces and address mapping work in the presence of virtual environments or intermediate schedulers present in such platforms.

%
\bibliographystyle{ACM-Reference-Format}
\bibliography{reference}

%
\end{document}